\documentstyle{article}

\begin{document}
\title{Supersymmetry Theory of Disordered Heteropolymers}

\author{A.I. Olemskoi, V.A. Brazhnyi\\
{\it Physical Electronics Department, Sumy State University}\\
{\it 2, Rimskii-Korsakov St., 244007 Sumy, UKRAINE }}

\maketitle

\begin{abstract}

The effective motion equation that describes the
different monomer alternation along the heteropolymer chain
is proposed.
On its basis the supersymmetry field scheme that allows to obtain the
equations for the structure factor and Green function is built up.
The memory and ergodicity breaking effects are investigated
depending on
the temperature and quenched disorder of the monomer alternation.
The phase diagram that determines the existence of the non-ergodic
and freezing states is provided.


\end{abstract}

\section{Introduction}

In recent years investigation of the
unusual behaviour of random heteropolymers with
a random sequence of different kinds of monomers
has attracted considerable attention (see \cite{3}-\cite{5} and
references therein).
With temperature decreasing such systems experience two
types of phase transitions
at which polymer either goes to freezing
state or has a microphase separation
that is inherent in  protein biomolecules type.

At folding process heteropolymer chooses out
of infinite set of possible states the  unique frozen
non-equilibrium but stationary state  with fixed
spatial conformation and monomer sequence (type of DNA).
The problem in considering of such transitions is to build up more
simple method of averaging thermodynamic values over
the quenched disorder.
Here, the methods of statistical physics first used in spin glass
investigation are taken as  basic ones \cite{6}.

Certain of polymers type of block copolymers, undergo the
so-called  microphase separation at which blocks of
different chemical composition segregate to microdomain structures to
gain a spatially inhomogeneous composition throughout the pattern.
The chemical  bonds between the blocks prevent the usual
macroscopic phase separation observed in homopolymer blends.
A variety of microphases,
such as one-dimensional lamellar structures,
hexagonal arrays of cylinders, body-centered-cubic
arrays of spheres and so on have been observed
for simple diblock copolymer melts.

At first, microphase separation had been studied
within mean-field approximation \cite{7} in the case of  A-B block
copolymers with arbitrary composition $f$.
It turns out that
 at $f\not= 0.5$ with temperature decrease the system
undergoes a sequence
of the microphase separations  of the first order,
where period
$2\pi/k_0$ of space structure is of order of block length
and does not depend on temperature.
In \cite{Braz} it was shown a
principal role of the fluctuations in phase transition.
By virtue of condition $k_0\not=
0$  the order parameter fluctuations give divergent
contribution to thermodynamic values, so that continuous
phase transition transforms to the first order weak transition
\cite{8}.

At passage to random heteropolymers both the microphase separation
and the freezing
remain, but here the space period acquires
 strong dependence on temperature \cite{3,9}.
According to field considerations \cite{10,11} the fluctuations
suppress both mentioned transitions in random copolymers,
but copolymer melts in disordered media can suffer these transformations.

Similar to spin glass \cite{6}, the main approach
of statistical mechanics of
random heteropolymers is the replica method
 \cite{3} that is not only pragmatic,
but very non-descriptive and artificial.
Apart from the replica  trick, the transfer matrix method
\cite{12}, kinetic  approach \cite{13} and others
(see \cite{14})
are used in random heteropolymers theory as well.
However, it is a well known
in theory of spin glass that within the framework of
Sherrington-Kirkpatrick model \cite{15} both replica method and
supersymmetry (SUSY) field approach  give identical results
\cite{16}.
On the other hand, the advantage of the SUSY field usage is
that  its components have
explicit physical meaning \cite{17}.

By recent moment, SUSY method in polymer theory had been proposed
by Vilgis \cite{18}, but did not obtain further development.
In this work we intend to complete this problem.
Our article is organized as follows.
In Section 2 we build up an effective motion equation
of disorder heteropolymer to describe the
different monomer alternation along the heteropolymer chain
\cite{20}.
The form of effective Hamiltonian of the problem under consideration
is obtained in Section 3 to avoid discrepancies of different methods.
Section 4 is devoted to consideration of a SUSY scheme \cite{21}
that is optimal for study of the disorder heteropolymer
in weak segregation limit \cite{22}.
In Section 5 a SUSY correlator is considered to examine
memory and non-ergodicity effects (Section 6).
Finally, Section 7 is devoted to discussion of obtained results.

\section{Effective motion equation}

Usually, the SUSY method is based on the dynamic
Langevin-type motion equation  \cite{21}.
But covalent bonds in the polymer chain
make such way inefficient because dynamic theory in polymers is
much more complicated, than statistical mechanics of usual many-body
systems \cite{23}.
Therefore, it is essential to
obtain the effective motion equation
instead of dynamic  one.

Let us initiate with directed polymer that is represented by
 the Gaussian
chain with probability $\Psi({\rm{\bf R}},N)$ to find the
end-point $N$ at coordinate ${\rm{\bf R}}$. As is known the
function $\Psi({\rm{\bf R}},N)$ obeys the Schrodinger-type
equation with imagine time $-{\rm i}N$ \cite{23}

\begin{equation}
\partial\Psi/\partial N=
\left (D\partial^2/\partial{\rm{\bf R}}^2-U({\rm{\bf R}},N)\right)\Psi,
\label{1}
\end{equation}
where number $N\gg 1$, $D\equiv b^2/6$ is effective diffusion coefficient
determined by the Kuhn segment length $b$, $U({\rm{\bf
R}},N)$ is external field.
In the limit $N\to\infty$ a solution of
Eq.(1) can be represented in the form of functional integral over
dependence ${\rm{\bf r}}(n)$ of chain coordinate on
number of internal monomers:

\begin{equation} \Psi({\rm{\bf
R}},N)=\int\exp(-S_{{\rm{\bf R}}N}\{{\rm{\bf
r}}(n)\}/2D)\delta{\rm{\bf r}}(n).
\label{2} \end{equation}
Here conventional action $S({\rm{\bf R}},N)\equiv S_{{\rm{\bf R}}
N}\{{\rm{\bf r}}(n)\}=\int_0^N L_0({\rm{\bf r}}(n)){\rm d}n$
corresponds to the fixed end-points ${\rm{\bf r}}(0)={\rm{\bf 0}}$,
and ${\rm{\bf r}}(N)={\rm{\bf R}}$ is  determined by Lagrangian
of the Euclidian field theory \cite{21,23}

\begin{equation} L_0={1\over
2}\left ({{\rm d}{\rm{\bf r}}(n)\over {\rm d}n}\right
)^2+2DU({\rm{\bf r}},n),
\label{3} \end{equation}
where the effective
kinetic energy, for which the continuum limit ${\rm{\bf
r}}(n+1)-{\rm{\bf r}}(n)\to{\rm d}{\rm{\bf r}}(n)/{\rm d}n$ is taken
into account, presents the covalent bonds between
monomers of polymer chain \cite{23}.
Inserting of ansatz (\ref{2}) into Eq.(\ref{1})
gives  the Jacobi-type equation

\begin{equation} {\partial
S\over\partial N}= D{\partial^2 S\over\partial{\rm{\bf R}}^2}-{1\over
2}\left ({\partial S\over\partial {\rm{\bf R}}}\right )^2+2DU.
\label{4} \end{equation}
After introducing generalized momentum
${\rm{\bf p}}\equiv\partial S/\partial{\rm{\bf R}}$ and total
derivative ${\rm d}{\rm{\bf p}}/{{\rm d}N}\equiv\partial{\rm{\bf
p}}/\partial N+({\rm{\bf p}}\partial/\partial{\rm{\bf R}}){\rm{\bf
p}}$, Eq.(4) takes the linear form of Burgers equation:

\begin{equation} {\rm d}{\rm{\bf
p}}/{\rm d}N=D\left (\partial^2{\rm{\bf p}}/\partial{\rm{\bf
R}}^2+2\partial U/\partial{\rm{\bf R}}\right ).  \label{5}
\end{equation}
The above pointed relation of
Eqs.(\ref{1})--(\ref{5}) is a well-known fact in
the theory of directed polymers,
stochastic growth and kinetic roughening phenomena (see \cite{24}).

The basic observation  for our purpose is that the
Schrodinger-type equation (\ref{1}) takes the form of the
Fokker-Planck equation \cite{25}

\begin{equation} {\partial P\over\partial
N}=\left (D{\partial^2\over\partial{\rm{\bf
R}}^2}-{\partial\over\partial{\rm{\bf R}}}{\rm{\bf F}}\right )P
\label{6} \end{equation}
for distribution probability

\begin{equation}
P({\rm{\bf R}},N)=\Psi({\rm{\bf R}},N)\exp\{-V({\rm{\bf R}})/2D\},
\label{7}
\end{equation}
which dependence on ${\rm{\bf R}}$ is determined by effective
potential

\begin{equation}
V\equiv -\int{\rm{\bf F}}{\rm d}{\rm{\bf R}}.
\label{8}
\end{equation}
Corresponding force ${\rm{\bf F}}$  relates with  the
initial potential $U$ in Eq.(\ref{1}) as follows:

\begin{equation}
U={1\over 4D}{\rm{\bf F}}^2+{1\over 2}{\partial{\rm{\bf
F}}\over\partial{\rm{\bf R}}}.  \label{9}
\end{equation}
As is well known \cite{26},
Eq.(\ref{6}) determines the probability
to realize solution  of corresponding Langevin-type equation

\begin{equation}
\partial{\rm{\bf
R}}/\partial N={\rm{\bf F}}({\rm{\bf R}},N)+{\rm{\bf \zeta}}(N)
\label{10}
\end{equation}
for stochastic variable ${\rm{\bf R}}={\rm{\bf R}}(N)$.
Here  the separated
stochastic force ${\rm{\bf\zeta}}$ is  fixed by the
white-noise conditions

\begin{equation}
\langle\zeta(N)\rangle=0,\qquad
\langle\zeta(N)\zeta(N')\rangle=2D\delta (N-N'), \label{11}
\end{equation}
where the angular brackets denote average with respect
to the distribution (\ref{7}).

In going from the above considered case of directed polymer to the
main object of our interest, the random heteropolymer AB, the
coordinate ${\rm{\bf R}}$ of end-point $N$ turns into a stochastic
Ising variable $\theta(n)$, where $\theta(n)=1$ if $n$-th segment
is of type A and $\theta(n)=-1$ otherwise.
Corresponding to the quenched disorder in fixed
sequence of different type segments, the law
$\theta(n)$ of monomer alternation along the chain
is described by master equation
type of that determines the Glauber dynamics \cite{27}.
Corresponding sequence correlator
$\overline{\sigma(n)\sigma(n')}$ for effective spin
$\sigma(n)\equiv\theta(n)-\overline{\theta(n)}$, being
deviation of microscopic value $\theta(n)$ from average
$\overline{\theta(n)}$, takes the form \cite{12}

\begin{eqnarray}
&&\overline{\sigma(n)\sigma(n')}=C_2\exp(-|n-n'|/l),\nonumber\\
&&C_2\equiv 4f(1-f),\qquad f\equiv(1/2)(1+\overline{\theta(n)}),
\label{12}
\end{eqnarray}
where overbar denotes the averaging over composition
(quenched disorder).
Here two sequence characteristics appear
being correlation length $l$ and fraction $f$ of type-A
monomers.

Stochastic variable $\sigma(n)$ that possesses of the
correlator (\ref{12}) with exponential form,
is governed by the effective motion equation

\begin{equation}
{\rm d}\sigma/{\rm d}n=-\sigma/l+s(n), \label{13} \end{equation}
where the stochastic source $s(n)$ is the white noise:

\begin{equation}
\overline{s(n)}=0,\qquad \overline{s(n)s(n')}=2C_2l^{-1}\delta(n-n').
\label{14}
\end{equation}
Relation between the microscopic value $\sigma(n)$ and stochastic
$\delta$-correlated variable $s(n)$ is given by equation

\begin{equation}
\sigma(n)=\int_0^n e^{-(n-m)/l}s(m){\rm d}m,
\label{15}
\end{equation}
that is the solution of Eq.(\ref{13}).
On the contrary to the colored noise
$\sigma(n)$, the white noise $s(n)$ possesses of the
Gauss distribution function

\begin{equation}
P\{s(n)\}=(4C_2\pi/l)^{-1/2}\exp\left\{-{l\over
4C_2}\int_0^N s^2(n){\rm d}n\right\},
\label{16} \end{equation}
that determines quenched disorder with intensity
$4C_2l^{-1}$.
Respectively, the local averaged  field

\begin{equation} \eta({\rm{\bf r}},n)\equiv (4C_2)^{-1/2}
\overline{\sigma(n)\delta({\rm{\bf r}}-{\rm{\bf r}}(n))}
\label{17} \end{equation}
represents the order parameter (here and below the
monomer volume takes to be equal unity).

Apart from usual terms type of that in Eq.(\ref{10}),
effective motion equation for the field (\ref{17}) must contain  the inhomogeneity
contribution $D\partial^2\eta/\partial{\rm{\bf
r}}^2$ (as the first term in r.h.s. of Eq.(\ref{4})) that takes
Fourier transform $-Dk^2\eta_{{\rm{\bf k}}}$.
As a result,
the motion equation for the Fourier transformation

\begin{equation}
\eta_{{\rm{\bf k}}}(n)=N^{-1/2}\int \eta({\rm{\bf r}},n)e^{-{\rm
i}{\rm{\bf k}}{\rm{\bf r}}}{\rm d}{\rm{\bf r}} \label{18}
\end{equation}
of the field (\ref{17}) takes the Langevin form:

\begin{equation}
\partial \eta_{{\rm{\bf k}}}/\partial
n=-(ak)^2 \eta_{{\rm{\bf k}}}- \partial
{\cal H}/\partial\eta^*_{{\rm{\bf k}}}+\zeta_{\rm{\bf k}}.
\label{19} \end{equation}
Here, as above, the continuum approximation
for effective time $n\gg 1$ is used,
the characteristic distance
$a\equiv D^{1/2}=6^{-1/2}b$ is
determined by the Kuhn segment length $b$, and
determination of effective force $f_{{\rm{\bf k}}}=-{\partial
{\cal H}/\partial\eta^*_{{\rm{\bf k}}}}$ (cf. Eq.(\ref{8})) are taken
into account.
The white noise $\zeta_{{\rm{\bf k}}}=\zeta_{{\rm{\bf
k}}}(n)$ is defined by conditions type of Eqs.(\ref{11}):

\begin{equation}
\langle\zeta_{{\rm{\bf k}}}\rangle=0,\qquad \langle\zeta_{{\rm{\bf
k}}}^*(n)\zeta_{{\rm{\bf k'}}}(n')\rangle=\delta_{{\rm{\bf
k}}{\rm{\bf k'}}}\delta(n-n'), \label{20} \end{equation}
where angular brackets denote average over thermal disorder.

\section{Effective Hamiltonian}

To obtain the effective Hamiltonian
${\cal H}\{\eta\}\equiv\Omega'\{m\}$ \cite{28}
in Eq.(\ref{19}) the thermodynamic
potential $\Omega'\{m\}$, being the average with
respect to both the conformation and the sequence scattering,
should be
determined as a function of the averaged order parameter

\begin{equation}
m({\rm{\bf r}})\equiv\sum_n\langle\eta({\rm{\bf
r}},n)\rangle=(4C_2)^{-1/2}\sum_n\overline{\langle\sigma(n)\delta
({\rm{\bf r}}-{\rm{\bf r}}(n))\rangle}.
\label{21} \end{equation}
With this aim, let us write the
partition function in the form of the functional integral \cite{21}

\begin{eqnarray}
&&\hspace{-2cm} Z=\int \delta m({\rm{\bf r}})\exp\left\{C_2\chi\int
m^2({\rm{\bf r}}){\rm d}{\rm{\bf r}}\right\}\times \nonumber\\
&&\times\overline{\left\langle\delta\left\{\sum_{n}\delta({\rm{\bf
r}}-{\rm{\bf
r}}(n))-1\right\}\delta\left\{\sum_{n}
(4C_2)^{-1/2}\sigma(n)\delta({\rm{\bf
r}}-{\rm{\bf r}}(n))-m({\rm{\bf r}})\right\}\right\rangle}.
\label{22}
\end{eqnarray}
Here $\chi >0$ is the composition Flory parameter, the first
$\delta$-function takes into account incompressibility condition, the
second one reduces to determination (\ref{21}) of order parameter
$m({\rm{\bf r}})$. Further, one follows to represent the
$\delta$-functions as functional Laplace expansions over auxiliary
fields $J_\rho$, $J_m$. Then the average expression in
Eq.(\ref{22}) takes the exponential form with the exponent
$\int(J_{\rho}+J_m m){\rm d}{\rm{\bf
r}}-\Omega\{J_{\rho},J_m\}$, where the last term, being
averaged over sequence and configuration sets, gets conventional free
energy at given fields $J_{\rho}$, $J_m$. The steady-state
magnitudes $\bar J_{\rho}$, $\bar J_m$ of these fields are
determined by conditions $\delta \Omega/\delta J_{\rho}=0$, $\delta
\Omega/ \delta J_m=-m$.  Inserting expression for $\bar
J_{\rho}$, $\bar J_m$ to functional $\Omega\{J_{\rho},J_m\}$
gets the thermodynamic potential defined by equation $Z=\int\delta
m_{{\rm{\bf k}}}\exp(-\Omega'\{m_{{\rm{\bf k}}}\})$, where (see more
detailed Refs.\cite{3,12,29})

\begin{eqnarray} &&\Omega'=\sum_{{\rm{\bf
k}}}\tau_{{\rm{\bf k}}}|m_{{\rm{\bf k}}}|^2+{1\over 2}\sum_{{\rm{\bf
k}}{\rm{\bf k}}'}w_{{\rm{\bf k}}{\rm{\bf k}}'}|m_{{\rm{\bf
k}}}|^2|m_{{\rm{\bf k}}'}|^2+\int v({\rm{\bf r}}){\rm d}{\rm{\bf r}},
\label{23}\\ &&\tau_{{\rm{\bf k}}}\equiv \tau+(ak)^2,\qquad
\tau\equiv l^{-1}-C_2\chi. \nonumber \end{eqnarray}
Here the
inhomogeneity contribution is taken into account in the quadratic term,
averaging procedure with respect to the quenched disorder (\ref{15})
results in appearance of $l^{-1}$-terms,
double ${\rm{\bf k}}$-dependence of the kernel
$w_{{\rm{\bf k}}{\rm{\bf k'}}}=4N^{-1}(la)^{-2}({\rm{\bf
k}}^2+{\rm{\bf k'}}^2)^{-1}$ is stipulated by averaging over the
distribution (\ref{2}).
It is worth mentioning that  expression (\ref{23}) agrees with
Refs.\cite{3,12}, whereas
in Refs.\cite{10,30}, where replica approach was used,
the second summand of Hamiltonian
(\ref{23})  with the opposite sign was obtained.
To correspond the
self-action effects the integrand in last term has the
usual expansion form \cite{3,12}

\begin{eqnarray}
&&v=-(\mu/3!)m^3+(\lambda/4!)m^4;\nonumber\\
&&\mu\equiv 12C_3C_2^{-1/2}l^{-1},\qquad
\lambda\equiv24(1+5C_3^2/C_2)l^{-1},
\label{24}\\
&&C_2\equiv 4f(1-f),\qquad
C_3\equiv|1-2f|.  \nonumber
\end{eqnarray} \\

As mentioned above, this method is based on the Eqs.(\ref{12}) that
allows to express quenched disorder correlator
within the transfer matrix approach.
If one uses the more popular
replica method, then the  field
$J_m$ and the order parameter $m$ must take a replica index $\alpha$
under which in the Hamiltonian (\ref{23}) the summation
from $1$ to $n\to 0$ should be made \cite{6}.
Then the quadratic contribution takes the form

\begin{equation}
{1\over 2}\sum_{{\bf
k}\alpha}A_{\alpha\alpha}({\bf k})
|m_\alpha ({\bf k})|^2+
{1\over 2}\sum_{{{\bf k}\atop\alpha\not=\beta}}
A_{\alpha\beta}({\bf k})
m_\alpha({\bf k})m_\beta(-{\bf k}),
\label{3.36}
\end{equation}\noindent\\
where in the limit $n\to 0$ for coincided replica indices is
$A_{\alpha\alpha}({\bf k})\to 2\tau_{\bf k}$.
As it was found out in spin glasses, the peculiarity of the systems
with quenched disorder is governed by the hierarchy of
the phase space  that is characterized by random overlapping
of the  different replicas \cite{6}.
Therefore overlapping parameter
$A_{\alpha\beta}({\bf k})$
in the second term  (\ref{3.36}) is the stochastic variable
under which the average should be made.
Let us take the corresponding distribution in the simplest
Gauss form

\begin{eqnarray}
{\cal P}\{A_{\alpha\beta} ({\bf k})\}\propto \exp\left\{ -{1\over 8}
\sum_{{\bf k}_1{\bf k}_2 \atop\alpha\ne\beta}  u_{{\bf k}_1{\bf
k}_2}^{-1} A_{\alpha\beta} ({\bf k}_1)A_{\alpha\beta} (-{\bf k}_2)
\right\} \label{3.37}
\end{eqnarray}\noindent\\
where dispersion
$ u_{{\bf k}_1{\bf k}_2}\equiv
\sigma^2(la)^{-2}N^{-1}({\bf k}_1^2+{\bf k}_2^2)^{-1}$
is given by parameter $\sigma$ (see \cite{10}).
Then after averaging of the partition function
$Z=\int Dm_{{\bf k}}\exp(-\Omega'\{m_{{\bf k}}\})$
the second term in Eq.(\ref{3.36}) takes the form

\begin{eqnarray}
-{1\over 2} \sum_{{\bf k}_1{\bf k}_2 \atop \alpha\ne\beta}
u_{{\bf
k}_1{\bf k}_2} m_{\alpha}({\bf k}_1) m_{\beta}(-{\bf
k}_1)m_{\alpha}(-{\bf k}_2) m_{\beta}({\bf k}_2). \label{3.38}
\end{eqnarray}\noindent\\
As a result the thermodynamic potential  (\ref{23})
in the replica form is

\begin{eqnarray}
&&\Omega'=\sum_{{\bf k}\alpha}\tau_{\bf k}|m_\alpha({\bf
k})|^2+{1\over 2}\sum_{{\bf k}_1{\bf k}_2\atop \alpha} w_{{\bf
k}_1{\bf k}_2}
|m_\alpha ({\bf k}_1)|^2|m_\alpha ({\bf
k}_2)|^2+ \nonumber\\
&&\label{3.39}\\
&&+\sum_\alpha\int v(m_\alpha){\rm d}{\bf r}-
{1\over 2} \sum_{{\bf k}_1{\bf k}_2\atop \alpha\ne\beta} u_{{\bf
k}_1{\bf k}_2} m_{\alpha}({\bf k}_1) m_{\beta}(-{\bf
k}_1)m_{\alpha}(-{\bf k}_2) m_{\beta}({\bf k}_2).
\nonumber \end{eqnarray}\\
The Eq.(\ref{3.39}) removes the above mentioned contradiction
in the choice of the effective Hamiltonian form:
the positive contribution of the second term
is governed by the  intra-replica interaction (see \cite{12,3}),
whereas negative contribution in \cite{30, 10} is determined
by replica overlapping.
The
distinction of mentioned terms makes itself evident in the fact that
the former leads to renormalization of the value  $\tau_{\bf k}$,
whereas the latter  corresponds to the memory and non-ergodic effects.

In order to carry out such renormalization (see \cite{28})
using of the mean-field approximation it needs to replace one of the
multiplier $|m_\alpha({\bf k})|^2$ in the second term of
Eq.(\ref{3.39}) by the bare Green function $G_{{\bf k}0}$
that corresponds to  $v=0$, $ u_{{\bf k}_1{\bf k}_2}=0$ and is
determined by equation

\begin{equation} G_{{\rm{\bf k}}0}^{-1}=r+2a^2(k-k_0)^2,
\label{25}
\end{equation}
independent on replica number $\alpha$.
To determine
the parameters $r$, $k_0$ one needs to substitute the
Eq.(\ref{25}) into the corresponding Dyson equation

\begin{equation} G_{{\rm{\bf k}}0}^{-1}=\tau_{{\rm{\bf
k}}}+\sum_{{\rm{\bf k'}}}w_{{\rm{\bf k}}{\rm{\bf k'}}} G_{{\rm{\bf
k'}}0}.  \label{26} \end{equation}
Then after integration over
wave vector ${\rm{\bf k'}}$ one obtains

\begin{equation}
r=\tau+(3/4\pi)l^{-2}(2r)^{-1/2},\qquad
k_0^{-1}=2\pi^{1/2}l(2r)^{1/4}a.
\label{27}
\end{equation}
According to the first equation the positive determined parameter $r$
increases smoothly with growth of bare parameter $\tau$
($r\sim\tau^{-2}$ at $\tau<0$, $|\tau|\gg 1$ and $r\sim\tau$ at
$\tau\gg 1$ (see Fig.1a)).
It means that without   self-action effect, random
heteropolymer is stable with respect to the microphase separation
\cite{8}.
The second equation (\ref{27}) means that space period
$\lambda\equiv 2\pi/k_0$ depends
on renormalized thermodynamic parameter $r$,
as it is inherent in random copolymers \cite{3}.
According to Fig.1b with increasing of
$\tau$  the value
$\lambda$ grows monotonously from $0$ to $\infty$.
In so doing the
greater the correlation length $l$, the faster change of
the period $\lambda$ near the point $\tau=0$.

The  final form of the effective Hamiltonian of
random heteropolymer   follows from the thermodynamic
potential (\ref{3.39}) renormalized under acting of
the fluctuations

\begin{eqnarray}
&&{\cal H}=\sum_{{\bf k}\alpha}\tau_{\bf k}|\eta_\alpha({\bf k})|^2
+ \sum_\alpha\int v(\eta_\alpha){\rm d}{\bf r}-
{1\over 2} \sum_{{\bf k}_1{\bf k}_2\atop \alpha\ne\beta}
u_{{\bf
k}_1{\bf k}_2} \eta_{\alpha}({\bf k}_1) \eta_{\beta}(-{\bf
k}_1)\eta_{\alpha}(-{\bf k}_2) \eta_{\beta}({\bf k}_2),
\label{28}
\\ &&r_{{\rm{\bf
k}}}\equiv r+2a^2(k-k_0)^2, \quad  u_{{\bf k}_1{\bf k}_2}\equiv
\sigma^2(la)^{-2}N^{-1}({\bf k}_1^2+{\bf k}_2^2)^{-1} .\nonumber
\end{eqnarray}

\noindent  Here the kernel
$v(\eta_\alpha)$ is determined by Eqs.(\ref{24}), where $m$ is
replaced by $\eta_\alpha$.

\section{Supersymmetric scheme}

To build a SUSY scheme on a basis of
effective motion equation (\ref{19})
let us introduce the generation functional \cite{21}

\begin{equation}
Z\{\eta_{\rm{\bf
k}}\}=\left\langle\delta\left ({\partial \eta_{\rm{\bf
k}}\over\partial n}+{\delta {\cal H}\over \delta \eta_{\rm{\bf
k}}^*}-\zeta_{\rm{\bf k}}\right )\det\left|{\delta\zeta_{\rm{\bf
k}}\over \delta \eta_{\rm{\bf k}}}\right|\right\rangle,\qquad
{\delta {\cal H}\over \delta \eta_{\rm{\bf k}}^*}\equiv{\partial
{\cal H}\over \partial \eta_{\rm{\bf
k}}^*}+2a^2(k-k_0)^2\eta_{\rm{\bf k}}, \label{33} \end{equation}
averaged over noise $\zeta_{{\rm{\bf k}}}(n)$, where
$\delta$-function considers the motion equation (19), the
determinant is Jacobian of transfer from $\zeta_{{\rm{\bf k}}}$ to
$\eta_{{\rm{\bf k}}}$. Then, the functional Laplace representation is
used for $\delta$-function that introduces a ghost field
$\varphi_{{\rm{\bf k}}}(n)$. To attach exponential form for the
determinant in Eq.(\ref{33}) Grassmann
conjugate fields $\psi_{{\rm{\bf k}}}(n)$, $\bar\psi_{{\rm{\bf
k}}}(n)$ should be used, that obey the conditions \cite{21}

\begin{eqnarray} &&
\{\psi,\psi\}=\{\psi,\bar\psi\}=\{\bar{\psi},\bar{\psi}\}=0,
\nonumber\\
&&\int\delta\psi=\int\delta\bar{\psi}=0,
\qquad\int\bar{\psi}\psi\delta^2\psi=1,\qquad
\delta^2\psi\equiv\delta\psi\delta\bar{\psi},
\label{34} \end{eqnarray}
where figure brackets denote anticommutator. Then, assuming that
the averaging over $\zeta_{\rm{\bf k}}(n)$
in Eq.(\ref{33}) is determined by Gauss distribution
with variance 1 (see Eqs.(\ref{20})), the standard form
of the partition function is obtained:

\begin{eqnarray}
&& Z\{\eta\}=\int
P\{\eta,\varphi;\psi,\bar{\psi}\}\delta\varphi\delta^2\psi,
\nonumber\\ &&
P\{\eta,\varphi;\psi,\bar{\psi}\}=\exp(-S\{\eta,\varphi;
\psi,\bar{\psi}\}),\qquad S=\int_0^N L{\rm d}n,
\label{35}\\
&& L=\int[(\varphi\dot
\eta-\bar{\psi}\dot\psi-\varphi^2/2)+({\cal
H}'\{\eta\}\varphi-\bar{\psi}{\cal H}''\{\eta\}\psi)]{\rm d}{\rm{\bf
r}}.  \nonumber \end{eqnarray}
Here point denotes derivative over
"time" $n$, the prime denotes the functional derivation with respect
to the field (\ref{17}), the last expression is taken in ${\rm{\bf
r}}$-representation.

The last expression in Eqs.(\ref{35}) takes the simplest form
\cite{22}

\begin{equation}
L=\int \Lambda (\Phi){\rm
d}^2\vartheta,\qquad \Lambda\equiv\sum_{\rm{\bf k}}
(\bar{{\cal D}}\Phi_{\rm{\bf
k}}^*)({\cal D}\Phi_{\rm{\bf k}})+{\cal H}\{\Phi_{\rm{\bf
k}}\}, \qquad {\rm d}^2\vartheta\equiv{\rm
d}\vartheta{\rm d}\bar{\vartheta}
\label{36}
\end{equation}
if one introduces SUSY generators

\begin{equation}
{\cal D}\equiv{\partial\over\partial\bar{\vartheta}}-
2\vartheta{\partial\over\partial n},
\qquad \bar{{\cal D}}\equiv{\partial\over\partial
\vartheta}
\label{37}
\end{equation}
and SUSY field

\begin{equation}
\Phi=\eta+\bar{\psi}\vartheta+\bar{\vartheta}\psi+
\bar{\vartheta}\vartheta\varphi,
\label{38}
\end{equation}
where Grassmann coordinates $\vartheta,\bar{\vartheta}$ obey the
same relations (\ref{34}) as for the fields $\psi,\bar{\psi}$.
Here functional ${\cal H}\{\Phi\}$ has the same form as the effective
Hamiltonian (\ref{28}), where order parameter $m_{{\bf k}\alpha}$ is
replaced by superfield $\Phi_{\bf k}(\vartheta)$, Eq.(\ref{38}).
In this case replica index $\alpha$ is removed by Grassmann variable
$\vartheta$ that is the formal reason of the replica and
supersymmetry methods identity.
The advantage of the latter method is in the Grassmann conditions
(\ref{34}) that correspond
to the limit $n\to 0$ in the replica
method.

According to consideration \cite{17,22},
the physical meaning of the components
of the SUSY field (\ref{38}) is as follow:
$\varphi$ is the most probable value of
fluctuations of the field conjugated to the order parameter $\eta$,
and combination $\bar{\psi}\psi$
determines density of sharp interphases.
So, using the
4-component SUSY field (\ref{38}) corresponds to the strong
segregation limit \cite{4,5}.
Further we shall consider more simple case of
the weak segregation limit where $\bar{\psi}\psi\equiv 0$.
Then, the SUSY field (\ref{38}) is reduced to the
2-component form

\begin{equation} \phi=\eta+{\theta}\varphi,
\label{39} \end{equation}
where self-conjugate nilpotent
variable ${\theta}\equiv\bar{\vartheta}\vartheta$ is introduced.
Respectively, Lagrangian (\ref{36}) takes the form

\begin{eqnarray}
&& L=\int\Lambda(\phi){\rm d}{\theta}, \qquad \Lambda\equiv
\sum_{\rm{\bf k}}\phi_{\rm{\bf k}}^*D\phi_{\rm{\bf k}}+
{\cal H}\{\phi_{\rm{\bf k}}\}; \nonumber\\
&& D=-{\partial\over\partial\theta}+\left
(1-2\theta{\partial\over\partial\theta}\right )
{\partial\over\partial n}.  \label{40} \end{eqnarray}
The motion equation
for the nilpotent field (\ref{39})
that corresponds to Lagrangian (\ref{40}) reads

\begin{equation} D\phi_{\rm{\bf k}}=-\delta
{\cal H}/\delta\phi_{\rm{\bf k}}^*. \label{41} \end{equation}
In the component form it leads to the equations for the order
parameter $\eta(n)$ and the amplitude of the more probable
fluctuation $\varphi(n)$ of the conjugate field  (see \cite{22}).

\section{Correlators}

Now let us consider SUSY correlator

\begin{equation}
C_{\rm{\bf k}}(n,{\theta};n',\theta ')\equiv
\left\langle\phi_{\rm{\bf k}}^*(n,\theta)\phi_{\rm{\bf
k}}(n',\theta ')\right\rangle. \label{42}
\end{equation}
Multiplying Eq.(\ref{41}) by value $\phi_{\rm{\bf
k}}^*$ and averaging  within zeroth approximation
($v=u=0$ in Eqs.(\ref{28})) one gets \cite{22}

\begin{equation}
C_{\nu{\rm{\bf k}}}^{(0)}(\theta,\theta ')={1+(r_{\rm{\bf
k}}-{\rm i}\nu)\theta+(r_{\rm{\bf k}}+{\rm i}\nu)\theta
'\over r_{\rm{\bf k}}^2+\nu^2}.  \label{43} \end{equation}
Here
conventional frequency $\nu$ denotes Fourier transformation over
"time" being the monomer number $n$, for example

\begin{equation}
C_\nu=\int_0^N C(n)e^{{\rm i}\nu n}{\rm d}n.\label{44}
\end{equation}
The expression (\ref{43}) has characteristic structure
with respect to combination
of the nilpotent variables $\theta$, $\theta '$, that is
inherent in, obviously, not only zeroth approximation but arbitrary
supercorrelator.
In this connection it is convenient to introduce
basis  supervectors

\begin{equation} A(\theta,\theta ')=\theta, \qquad
B(\theta,\theta ')=\theta ',\qquad T(\theta,\theta ')=1, \label{45}
\end{equation}
that have functional production

\begin{equation}
X(\theta,\theta ')=\int Y(\theta,\theta '')Z(\theta '',\theta){\rm
d}\theta '' \label{46} \end{equation}
for any vectors ${\bf X}$, ${\bf Y}$, ${\bf Z}$.
It is easy to see, that the basis  SUSY vectors (\ref{45})
obey the following multiplication rules: ${\bf A}^2={\bf A}$,
${\bf B}^2={\bf B}$, ${\bf B}{\bf T}={\bf T}$, ${\bf T}{\bf
A}={\bf T}$, other products are zero.
Because the set of
vectors ${\bf A}$, ${\bf B}$, ${\bf T}$ is complete, it
is convenient to expand any SUSY correlator over this basis:

\begin{equation}
{\bf C}=G_-{\bf A}+G_+{\bf B}+S{\bf T}. \label{47}
\end{equation}
Here and below subscripts ${\rm{\bf k}}$, $\nu$ are suppressed for
brevity.
Using Eqs.(\ref{39}), (\ref{42}), one obtains for coefficients
of the expansion (\ref{47})

\begin{equation}
G_-=\langle\eta\varphi^*\rangle,\qquad
G_+=\langle\eta^*\varphi\rangle,\qquad S=\langle|\eta|^2\rangle.
\label{48} \end{equation}
So, $G_{\pm}$ represent advanced and
retarded Green functions and $S$ is the structure factor.
In accordance with Eqs.(\ref{43}), (\ref{45}), (\ref{47}),
these functions
within zeroth approximation take the form

\begin{equation}
G_\pm^{(0)}=(r\pm{\rm i}\nu)^{-1}, \qquad
S^{(0)}=G_+^{(0)}G_-^{(0)}=(r^2+\nu^2)^{-1}. \label{49}
\end{equation}

The Dyson equation for SUSY correlator (\ref{42})
is as follows \cite{19,22}

\begin{equation}
{\bf C}^{-1}=({\bf C}^{(0)})^{-1}-u{\bf C}-{\bf \Sigma}.
\label{50}
\end{equation}
Here ${\bf\Sigma}$ is self-energy function,
$u=\sigma^2 (2N)^{-1}(lak_0)^{-2}$ is the typical value
of the  interreplica overlapping $u_{\bf kk'}$ at ${\bf k}=
{\bf k'}={\bf k}_0$.
By analogy with
the expansion (\ref{47})  the self-energy superfunction
${\bf \Sigma}$ that describes the self-action effects is

\begin{equation} {\bf \Sigma}=\Sigma_-{\bf A}+\Sigma_+{\bf
B}+\Sigma{\bf T}. \label{51} \end{equation}
Then using  Eqs.(\ref{49}),
the Dyson equations for component $G_\pm$, $S$ take
the form

\def\theequation{\arabic{equation}{a}}
\setcounter{equation}{51} \begin{equation}
G_\pm^{-1}+uG_\pm=(r\pm{\rm i}\nu)-\Sigma_\pm, \label{52a}
\end{equation} \def\theequation{\arabic{equation}{b}}
\setcounter{equation}{51} \begin{equation} S=(1+2\pi
C_2l^{-1}\delta(\nu)+\Sigma)G_+G_-(1-uG_+G_-)^{-1}.  \label{52b}
\end{equation}
\def\theequation{\arabic{equation}}\setcounter{equation}{52}

\noindent Here $\delta$-term is  caused by quenched disorder.

To complete the system (52) it needs to express  the components
of the self-energy superfunction by supercorrelators.
Using
the SUSY perturbation theory with accounting of the cubic and
quartic anharmonicities (\ref{24}) for matrix
element of the self-energy function one  obtains  \cite{21}

\begin{equation}
\Sigma(z,z')=
{\mu^2\over 2!}\left (C(z,z')\right )^2
+{\lambda^2\over 3!}\left (C(z,z')\right )^3.
\label{53} \end{equation}
Here $z\equiv\{{\rm{\bf r}},n,\theta\}$ denotes the set of coordinate
${\rm{\bf r}}$, "time" $n$, and nilpotent coordinate $\theta$.
It is important that the multiplication rule in
Eq.(\ref{53}) is not the same as for the functional product
(\ref{46}).
Here the ordinary product should be used
according to the following multiplication rules \cite{16,22}:
$AT=TA=A$, $BT=TB=B$, $T^2=T$, other products are zero.
Then coefficients of the SUSY expansion (\ref{51})
take the form

\def\theequation{\arabic{equation}{a}} \setcounter{equation}{53}
\begin{equation}
\Sigma_\pm(n)=
\left(\mu^2+{\lambda^2\over 2}S(n)\right)S(n)G_\pm(n), \label{54a}
\end{equation}
\def\theequation{\arabic{equation}{b}} \setcounter{equation}{53}
\begin{equation}
\Sigma(n)={1\over 2}\left(\mu^2+{\lambda^2\over
3}S(n)\right)S^2(n),\label{54b} \end{equation}
\def\theequation{\arabic{equation}}\setcounter{equation}{54}

\noindent where ${\rm{\bf r}}$-representation  for
macroscopically homogeneous system is used.

\section{Memory and non-ergodicity effects}

Following to Edwards and Anderson \cite{31}
let us introduce the composition  memory parameter
$q\equiv\langle\eta(n=N)\eta(n=0)\rangle$,
which value determines the long-range correlation
in the different monomer alternation
along the whole heteropolymer chain.
Moreover, the non-ergodicity parameter
$\Delta\equiv g_0-g$ that is difference between the isothermal
susceptibility $g_0\equiv G_-(\nu=0)$ and thermodynamic value
$g\equiv G_-(\nu\to 0)$ is used.
Then, the main correlators acquire the elongated form:

\begin{equation}
G_\pm(\nu)=\Delta+G_{\pm 0}(\nu),\qquad S(n)=q+S_0(n), \label{55}
\end{equation}
where index $0$ denotes the components corresponding to ergodic
system without memory.
Substitution of Eqs.(\ref{55}) into Eqs.(54) gives

\def\theequation{\arabic{equation}{a}} \setcounter{equation}{55}
\begin{eqnarray}
&&\Sigma_\pm(n)=\left (\mu^2+{\lambda^2\over 2}q\right )q\left
(\Delta+G_{\pm 0}(n)\right )+\Sigma_{\pm 0}(n),
\nonumber\\
&&\Sigma_{\pm 0}(n)\equiv(\mu^2+\lambda^2q)S_0(n)G_{\pm 0}(n)
+{\lambda^2\over 2}S_0^2(n)G_{\pm 0}(n); \label{56a}
\end{eqnarray}
\def\theequation{\arabic{equation}{b}} \setcounter{equation}{55}
\begin{eqnarray}
&&\Sigma(n)={1\over 2}\left (\mu^2+{\lambda^2\over 3}q\right
)q^2+\left (\mu^2+{\lambda^2\over 2}q\right )qS_0(n)+\Sigma_0(n),
\nonumber\\
&&\Sigma_0(n)\equiv{1\over
2}(\mu^2+\lambda^2q)S_0^2(n)+{\lambda^2\over 6}S_0^3(n).
\label{56b}
\end{eqnarray}
\def\theequation{\arabic{equation}} \setcounter{equation}{56}

\noindent Here the terms, being non-linear with respect to
correlators $G_{\pm 0}$, $S_0$, are included
into summands $\Sigma_{\pm
0}$, $\Sigma_0$, the terms
that contain production $S_0\Delta\approx 0$ have been
dropped and, finally, one separates out the first summands that
disappear when parameter $q$ goes to zero.
It is
characteristic that to determine the self-energy functions (56)
the "time"-representation was used, whereas the Dyson equations
(52) require "frequency" Fourier transformations of non-linear
expressions (56). To avoid this difficulty
the fluctuation-dissipation theorem is used \cite{22}

\def\theequation{\arabic{equation}{a}} \setcounter{equation}{56}
\begin{equation}
S_0(n\to 0)=G_{\pm 0}(\nu\to 0)\equiv g,
\label{57a}
\end{equation}
\def\theequation{\arabic{equation}{b}} \setcounter{equation}{56}
\begin{equation}
\Sigma_{\pm 0}(\nu\to 0)=\Sigma_0(n\to 0)\equiv{1\over
2}(\mu^2+\lambda^2q)g^2+{\lambda^2\over 6}g^3,
\label{57b} \end{equation}
\def\theequation{\arabic{equation}}\setcounter{equation}{57}

\noindent
where  Eq.(\ref{56b}) is taken into account in the last equation.

Inserting Fourier transformations of Eqs.(\ref{55}), (56) to
Dyson equation (\ref{52b}), within the $\nu$-representation
one obtains

\begin{equation}
q_0\left[1-ug_0^2-{1\over 2}\left(\mu^2+{\lambda^2\over
3}q_0\right)q_0g_0^2\right]=C_2l^{-1}g_0^2,
\label{58} \end{equation}
\def\theequation{\arabic{equation}{a}} \setcounter{equation}{57}
\begin{equation}
S_0={{(1+\Sigma_0)G_+G_-}\over{1-[u+(\mu^2+\lambda^2q/2)q]G_+G_-}}.
\label{58a}
\end{equation}
\def\theequation{\arabic{equation}}\setcounter{equation}{58}

\noindent The first of these equations corresponds to $\delta$-term
being caused by memory effects, the second one - to "frequency"
$\nu\not=0$. At $\nu\to 0$ the characteristic production is
$G_+G_-\to g^2$, so that the pole of structure factor (\ref{58a})

\begin{equation}
u+\left(\mu^2+{\lambda^2\over 2}q\right)q=g^{-2}
\label{59} \end{equation}
determines the point of ergodicity breaking.
By analogy,
substituting of Eqs.(\ref{56a}), (\ref{57b}) into the Dyson equation
(\ref{52a})  and taking into account $g\equiv G_-(\nu\to 0)$
for microscopic susceptibility and memory parameter we have

\begin{equation}
1-rg+ug^2+{\mu^2\over 2}g\left[(g+q)^2-q^2\right]+{\lambda^2\over
6}g\left[(g+q)^3-q^3\right]=0.
\label{60} \end{equation}
The thermodynamic behaviour of random heteropolymer in the
vicinity of the ergodicity breaking point is described  by the system
of equations (\ref{58})-(\ref{60}).
By analogy with
spin glass \cite{6}, Eqs.(\ref{58}), (\ref{60}) play role of
Sherrington-Kirkpatrick equations, and Eq.(\ref{59}) determines the
point of de Almeida-Thouless instability.
On further  analyzing
of these equations one should distinguish  macro- and
microscopic values $q_0$, $q$ of the memory parameter and
corresponding susceptibilities $g_0$, $g$.
The peculiarity of such
a hierarchy is that microscopic values, which conform to the limit $\nu
\to 0$, are the usual thermodynamic parameters and depend on
temperature (Flory parameter $\chi$).
The macroscopic values $q_0$, $g_0$ conform to the point $\nu =0$ and
only depend on quenched disorder parameter $l$.
In the non-ergodic area
macroscopic quantities take the values at the point of ergodicity
breaking.

Because the system of three equations (\ref{58})--(\ref{60})
is insufficient for determination of four quantities
$q_0$, $q$, $g_0$, $g$,
it  must be completed by equation

\begin{equation}
q\left[1-ug^2-{1\over 2}\left(\mu^2+{\lambda^2\over
3}q\right)qg^2\right]=C_2l^{-1}g^2,
\label{61}
\end{equation}
that is microscopic analogue of the equation (\ref{58})
obtained as a result of separating out of the singular $\delta$-terms
 for the structure factor in the Dyson equation.
As is known from the spin glass theory \cite{6},
the hierarchy of such singularities, which correspond to
a set of infinitely decreasing "frequencies" $\nu \to 0$,
is a key point of the non-ergodic systems.
In our case, the equations (\ref{58}), (\ref{61})
correspond to the point $\nu=0$ and a minimal of these "frequencies"
respectively.
In turn, at $\nu=0$ the equation (\ref{59}) reads:

\begin{equation}
u+\left(\mu^2+{\lambda^2\over 2}q_0\right)q_0=g_0^{-2}.
\label{62}
\end{equation}
Eqs.(\ref{58}), (\ref{60})--(\ref{62}) are the complete system
for determination of the quantities $q_0$, $q$, $g_0$, $g$.

\section{Discussion}

According to Eqs.(\ref{58}), (\ref{62}) the macroscopic memory
parameter is given by cubic equation

\begin{equation}
(\mu^2/2+\lambda^2q_0/3)q_0^2=C_2l^{-1}.
\label{63} \end{equation}
The characteristic form of the dependence of the value $q_0$
on intensity  of quenched disorder $l$ are depicted in Fig.2.
For copolymers close to symmetric composition $f=0.5$ ($C_3\ll
C_2$) the first term in brackets of Eq.(\ref{63}) is negligible, and
the dependence $q_0\propto l^{1/3}$ takes place. In
the opposite case of
dilute copolymer, where $f\ll 1$ ($C_2\ll C_3$), one obtains the
dependence $q_0\propto f~l^{1/2}$.

Simultaneous solution of Eqs.(\ref{58}), (\ref{60}) and (\ref{62})
gives the point of ergodicity breaking $\chi_0$,
which dependence on correlation length $l$
is depicted in Fig.3 (solid curve).
A characteristic of this phenomenon is that non-zeroth value of
$\chi_0$ appears above a critical value of correlation length $l$,
and with further growth of $l$ dependence reaches its maximum and
then  monotonously falls down.
In so doing, ergodic region is located under curve $\chi_0(l)$
and contracts  with correlation length growth.
Condition ${\rm d} g/{\rm d}\chi =-\infty$ added by Eqs.(\ref{60}),
(\ref{61}) gives the equation

\begin{equation} u+\mu^2(g_f+q)+{\lambda^2\over 2}(g_f+q)^2=
g_f^{-2},
\label{64} \end{equation}
that defines $\chi_f$ value of Flory parameter in the
freezing point, under which the microscopic susceptibility
$g$ takes zeroth value (see Fig.4).
Corresponding dependence  $\chi_f(l)$ on the correlation length $l$
is depicted in Fig.3 (thin curves).
A characteristic is that the dependence $\chi_f(l)$ is
under ergodicity
breaking curve $\chi_0 (l)$ and has the same form.
The influence of the composition on the values of  $\chi_0$
and $\chi_f$ is shown in Fig.3a.
As we recede from the composition $f=0.5$,
the growth of the above mentioned parameters is observed.
The more complicated behaviour is realized at the growth of the
interreplica overlapping  $\sigma$ (see Fig.3b).
At small magnitudes of the correlation length $l$,
 the growth of both $\chi_0$ and $\chi_f$
values is observed with $\sigma$ increasing,
whereas at large magnitudes $l$,
parameters $\chi_0$ and $\chi_f$ decrease.

Dependencies of the macroscopic $g_0$ and microscopic $g$
susceptibilities on the parameter $\chi$ are depicted in Fig.4.
Under point of  ergodicity breaking $\chi_0$, these susceptibilities
and corresponding  memory parameters $q$, $q_0$ coincide.
The dependence $g(\chi)$ has
a cut-off at the point $\chi=\chi_f$
(below it, susceptibility $g$ takes zeroth value
that  corresponds to the freezing state).
Above the ergodicity breaking point $\chi_0$,
the macroscopic susceptibility $g_0$ is constant and
the microscopic one has smooth decrease
(latter can be obtained by simultaneous solution
of Eqs.(\ref{60}), (\ref{61})).
According to Fig.4a, with moving away from the composition $f=0.5$,
the values of the susceptibilities $g_0$, $g$
in the freezing point and in the point of ergodicity breaking
decrease but corresponding values of
parameters $\chi_0$ and $\chi_f$ increase.
The dependence on the correlation length $l$
is shown in Fig.4b: with increasing of $l$ ergodic area contracts
as it should be.
The influence of the interreplica  overlapping is shown in Fig.4c.
With increasing of the corresponding parameter  $\sigma$, the values
of the susceptibilities $g(\chi)$ and $g_0(\chi)$ decrease and hence
the interreplica  overlapping  prevents the
heteropolymer freezing.

The influence of the thermodynamic parameter $\chi$ on the
microscopic memory parameter $q$ is depicted in Fig.5.
The absence of memory below $\chi_f$
is the characteristic feature of the freezing region.
Non-zeroth value
of $q$ appears in the freezing point $\chi_f$ and with further
growth of the parameter $\chi$, the memory parameter monotonously increases.
The step-like behaviour of the parameter $q$ is
inherent in the first order transition.
Evidently, the
physical reason of the mentioned  behaviour is the fluctuation
contribution into the thermodynamic potential of the heteropolymer.
According to Fig.5a, with moving away from the composition $f=0.5$,
the dependence $q(\chi)$ becomes more slight.
On the
contrary, the growing of the  correlation length $l$ results in
more abrupt growth of the memory parameter (see Fig.5b).
At last, one can see from Fig.5c that influence of the
interreplica overlapping  $\sigma$ above and below of the ergodicity
breaking point occurs to be opposite.

According to the Fig.6, the non-ergodicity parameter
$\Delta (\chi)$  grows monotonously with  increasing $\chi$ from
the   ergodicity breaking point $\chi_0$.
The deflection from the composition $f=0.5$, the decrease
of the correlation length $l$ and the growth of the interreplica
overlapping $\sigma$ result in  weakening of the non-ergodicity effects
(see Fig.6).

In the search for a new polymers with predetermined features,
the  phase diagram  plays the basic
role  that defines possible thermodynamic states
at different values of the Flory parameter $\chi$ and
composition $f$.
According to Fig.7, such diagram has a concave form
for both the freezing point $\chi_f(f)$ and the point
of ergodicity breaking $\chi_0(f)$.
A region of the large $\chi$ adjacent
to the composition $f=0.5$ corresponds to the non-ergodic
unfreezing state. With decreasing of the values $\chi$ and $|f-1/2|$,
at first the system goes to the ergodic state  and  then it freezes.
The comparison of the Fig.7a and Fig.7b shows that the increasing of
the correlation length $l$ results in expansion of the unfreezing
and non-ergodic phases.
On the contrary, from Fig.7a and Fig.7c one can see that interreplica
overlapping leads to its contraction.

\newpage

\section*{Captions}

\begin{description}

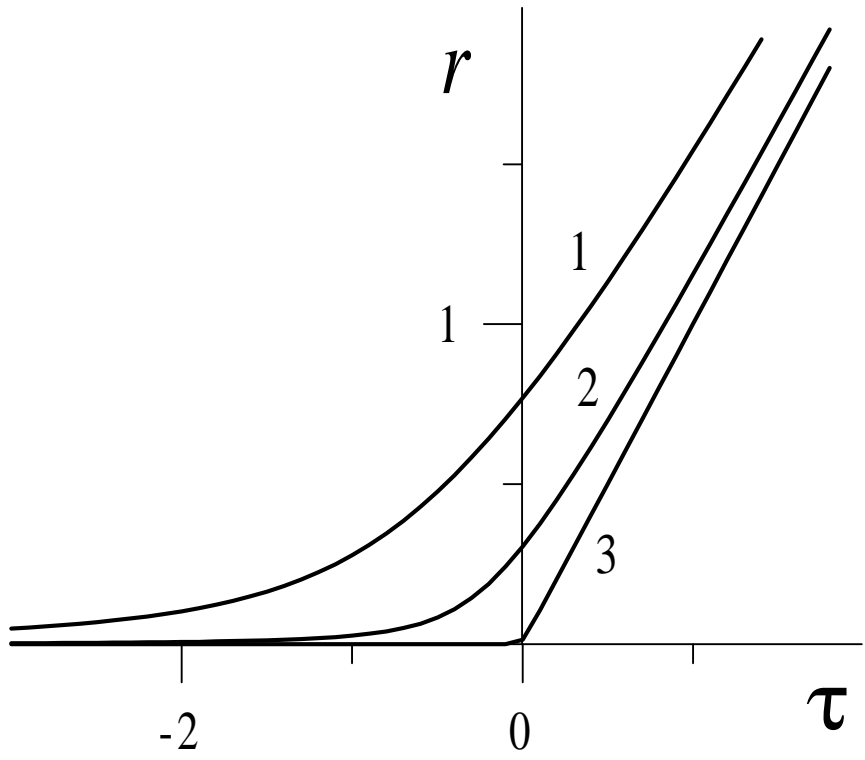
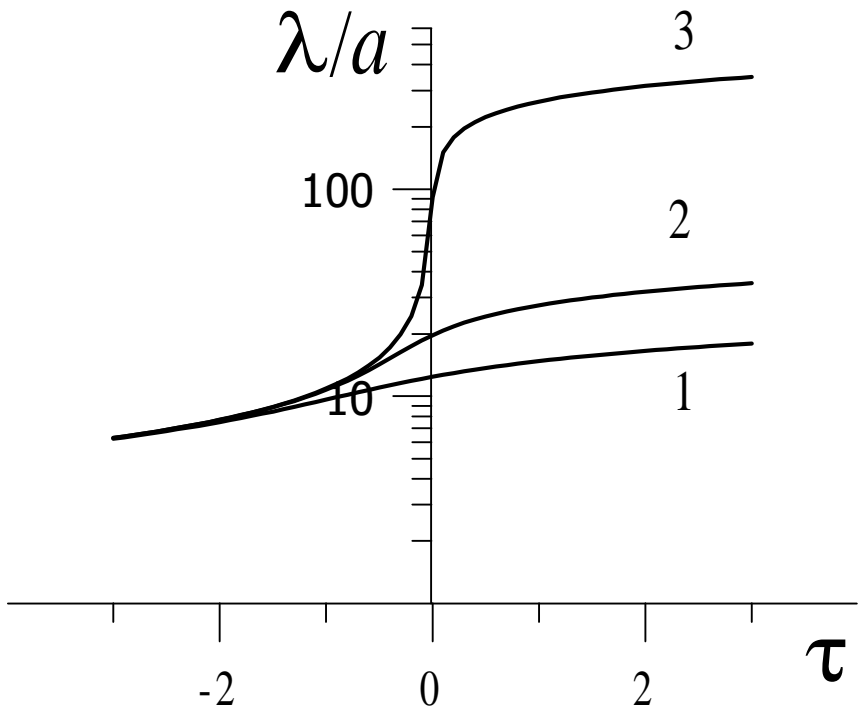
\item{Fig.1} Dependence of the parameter $r$ (a)
and  the period $\lambda$ (b) on  $\tau$ at
different values of the correlation
length $l$ (curves 1, 2, 3 correspond to $l=0.5; 1; 10$).

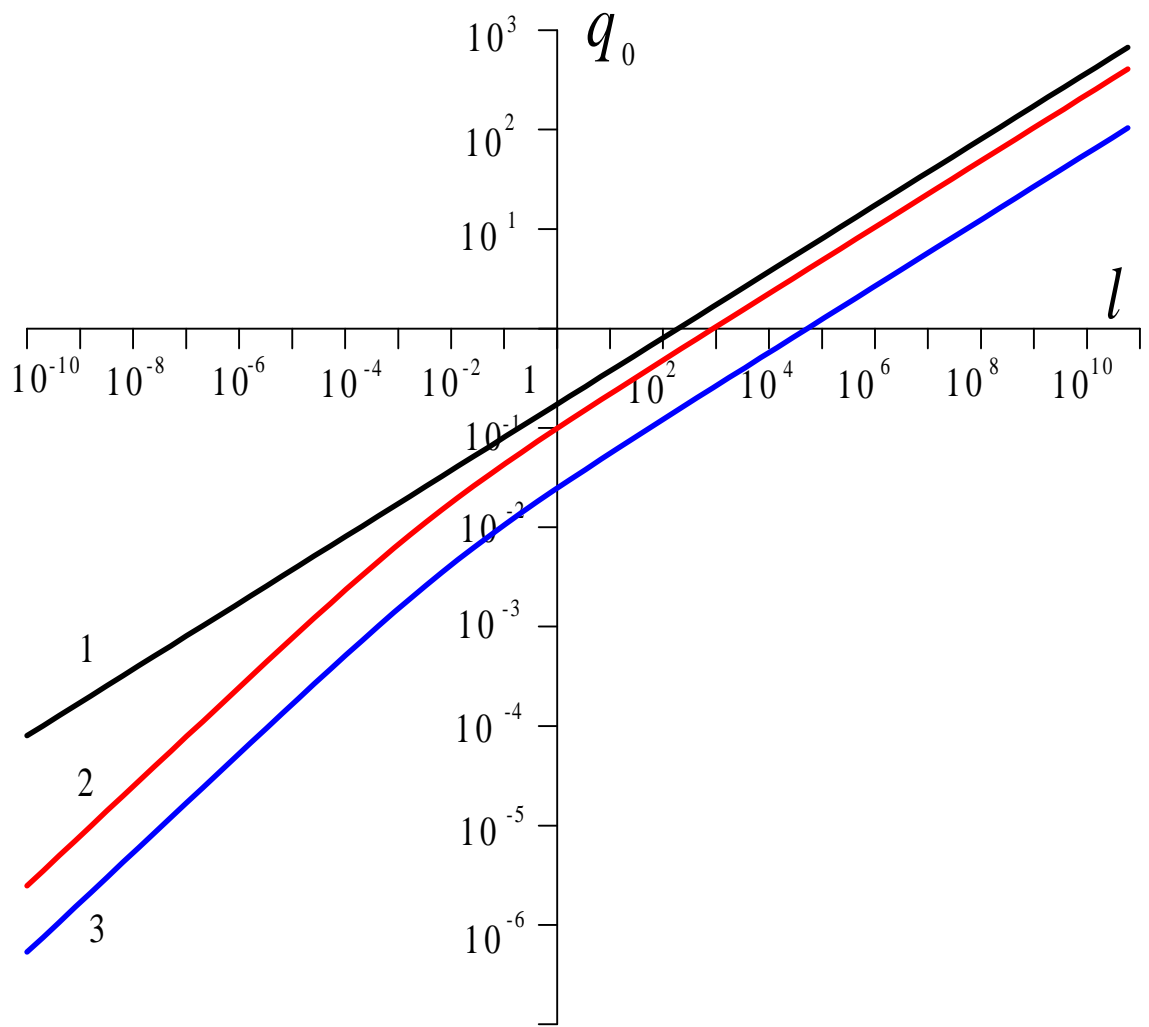
\item{Fig.2} Dependence  of the macroscopic
memory parameter  $q_0$ on the value of the correlation length $l$
(curves 1, 2, 3 correspond to $f=0.5; 0.3; 0.1$).

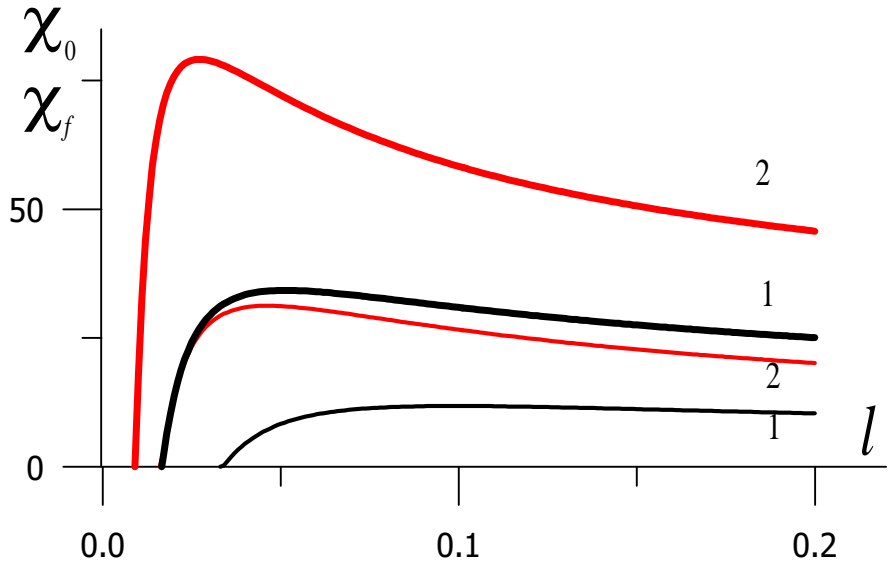
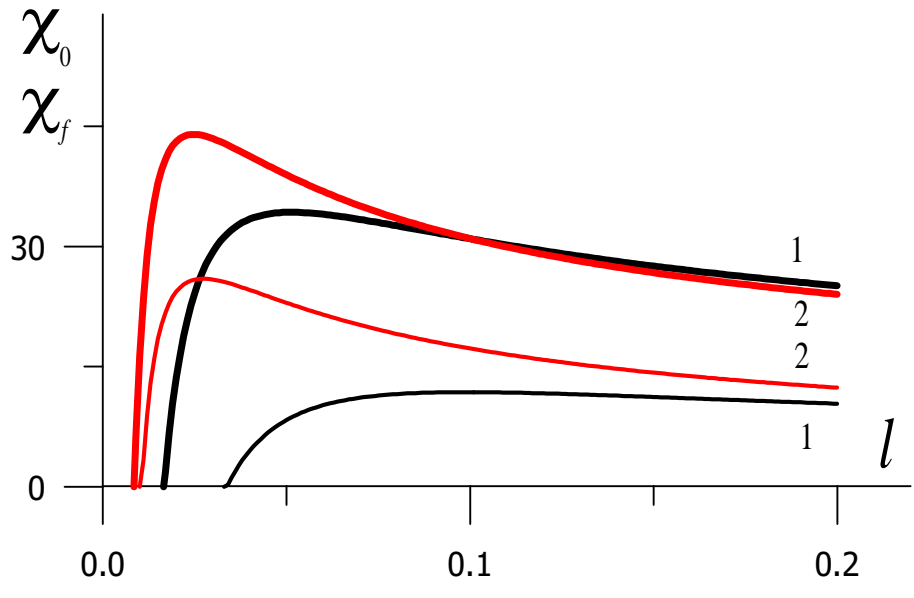
\item{Fig.3} Dependence  of the characteristic values of the Flory
parameter in the point of ergodicity breaking $\chi_0$
(thick curve)  and in the freezing point $\chi_f$ (thin curve) on
the correlation length $l$:
a) at $\sigma=0$ and different values of the composition $f$
(curves 1, 2 correspond to $f=0.5; 0.3$);
b) at $f=0.5$ and different values of
the interreplica overlapping $\sigma$
(curves 1, 2 correspond to $\sigma=0; 2$).

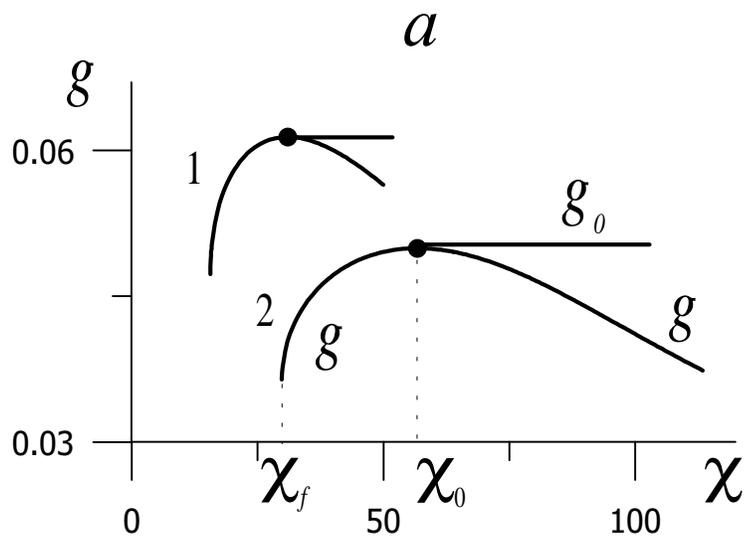
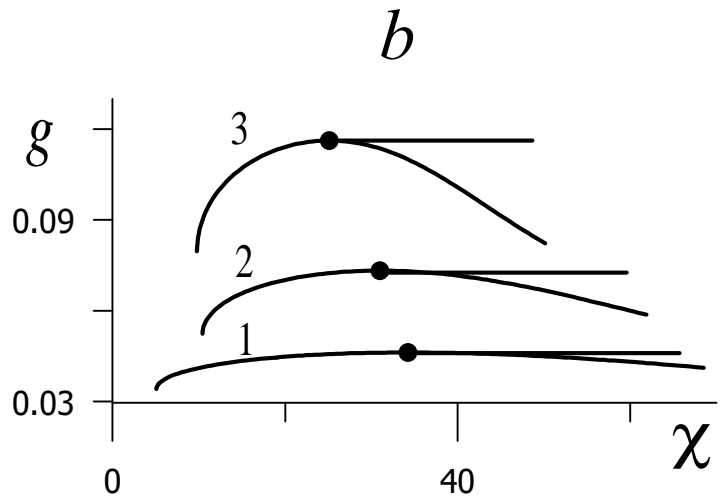
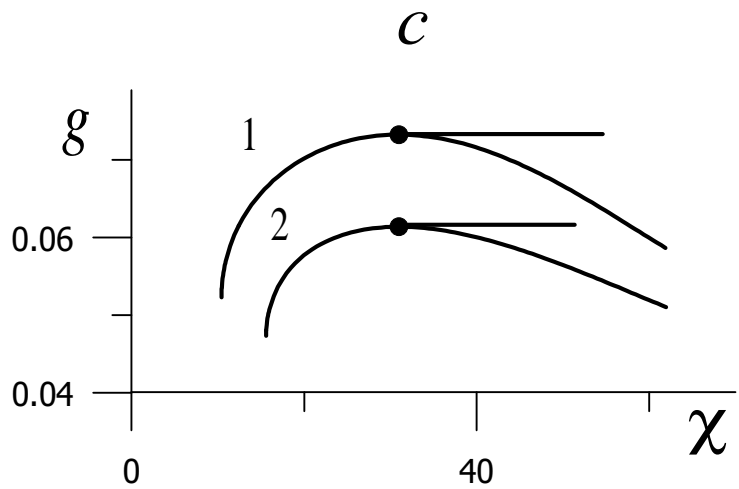
\item{Fig.4}  Dependence of the
microscopic  $g$ and macroscopic
$g_0$ susceptibilities on the parameter  $\chi$:
a) at  $\sigma=0$, $l=0.1$ and
different values of the composition $f$
(curves 1, 2 correspond to $f=0.5; 0.3$);
b) at $f=0.5$, $\sigma=0$
and different values of
the correlation length $l$
(curves 1, 2, 3 correspond to $l=0.05; 0.1; 0.2$);
c) at $f=0.5$, $l=0.1$ and different values of the interreplica
overlapping parameter  $\sigma$
(curves 1, 2 correspond to $\sigma=0; 2$).

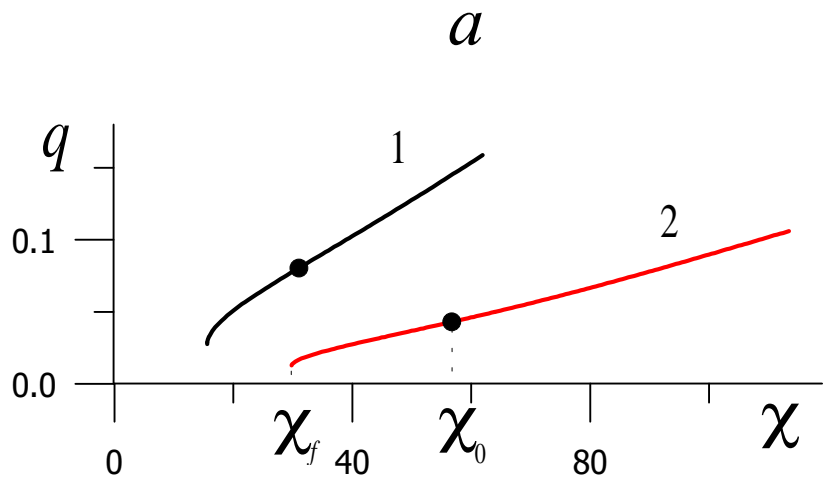
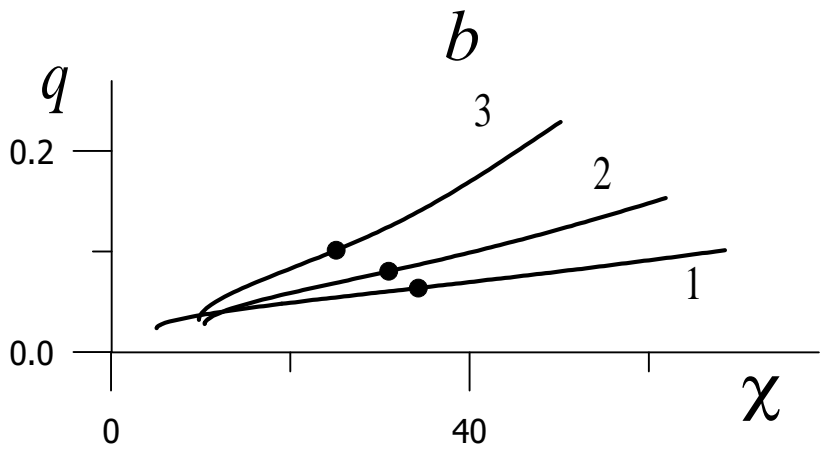
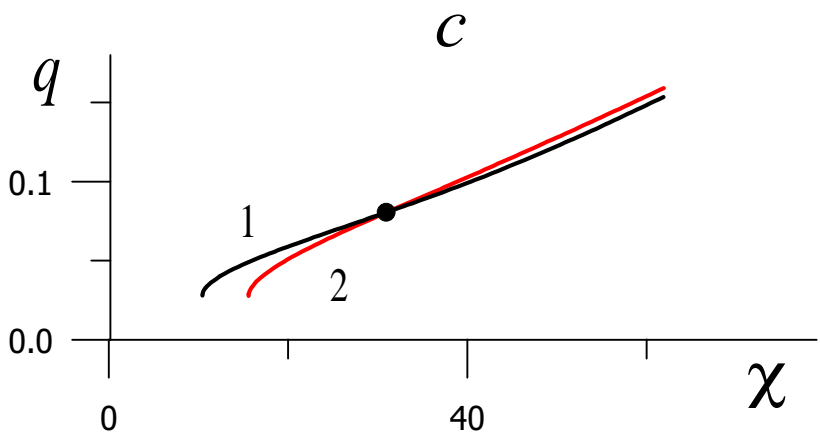
\item{Fig.5}  Dependence of the microscopic memory
parameter  $q$ on the Flory parameter $\chi$:
a) at $\sigma=0$, $l=0.1$ and
different values of the composition $f$
(curves 1, 2 correspond to $f=0.5; 0.3$);
b) at  $f=0.5$, $\sigma=0$ and different values of the correlation
length $l$
(curves 1, 2, 3 correspond to $l=0.05; 0.1; 0.2$);
c) at $f=0.5$, $l=0.1$ and different  values of the interreplica
overlapping parameter $\sigma$
(curves 1, 2 correspond to $\sigma=0; 2$).

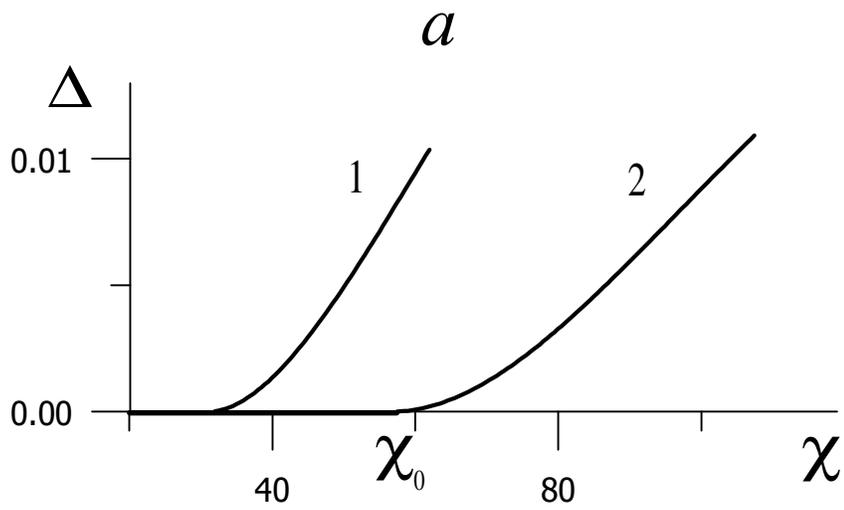
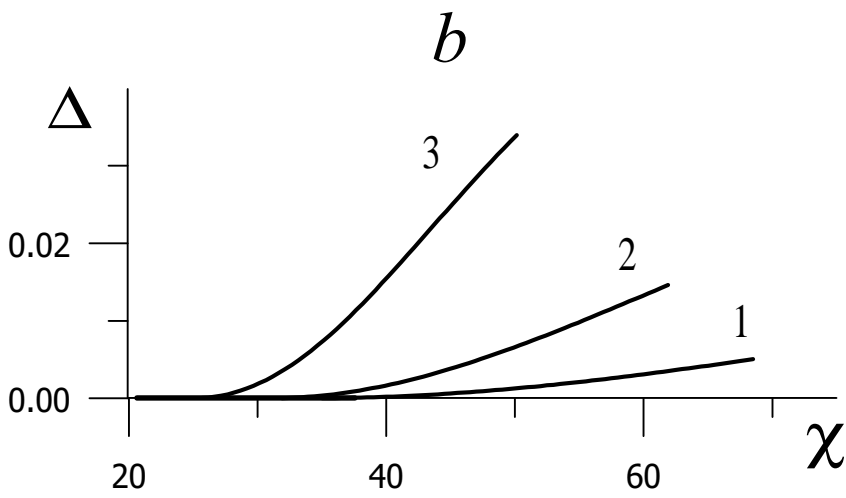
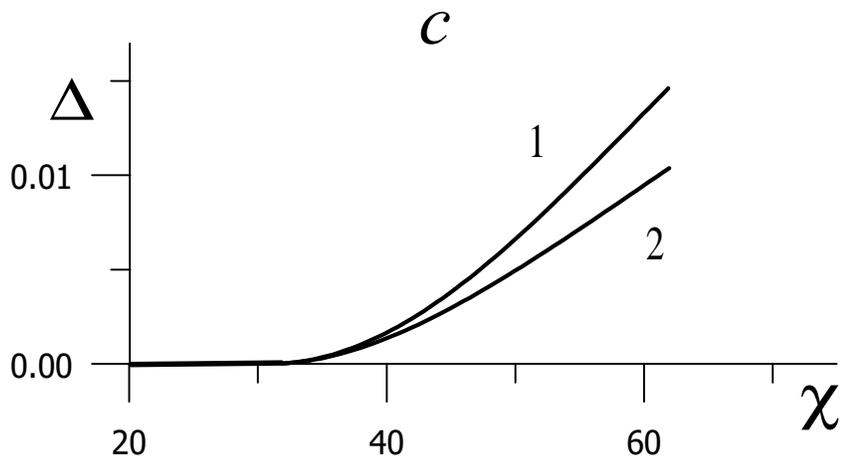
\item{Fig.6}  Dependence of the non-ergodicity parameter
 $\Delta$ on the parameter $\chi$:
a) at $\sigma=0$, $l=0.1$ and
different values of the composition $f$
(curves 1, 2 correspond to $f=0.5; 0.3$);
b) at  $f=0.5$, $\sigma=0$ and different values of the correlation
length $l$
(curves 1, 2, 3 correspond to $l=0.05; 0.1; 0.2$);
c) at $f=0.5$, $l=0.1$ and different values  of the
parameter $\sigma$
(curves 1, 2 correspond to $\sigma=0; 2$).

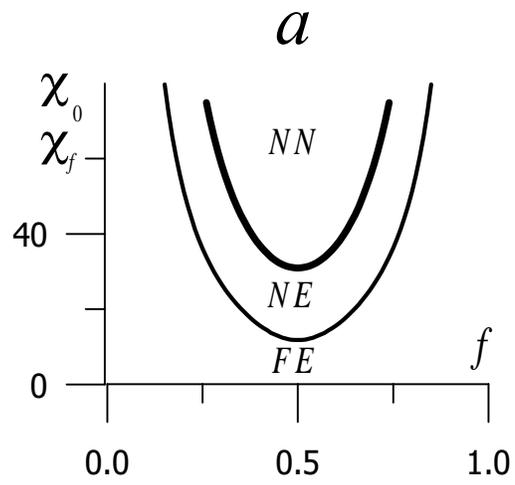
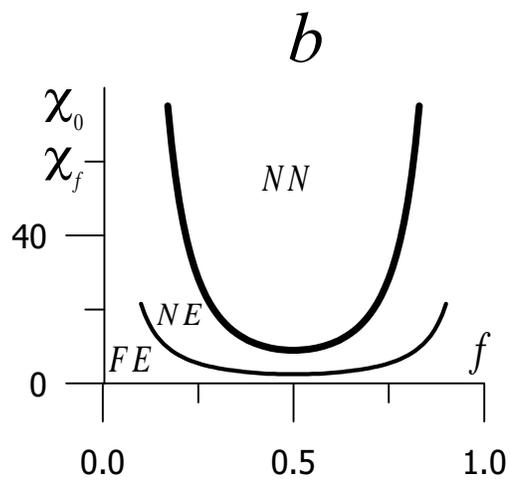
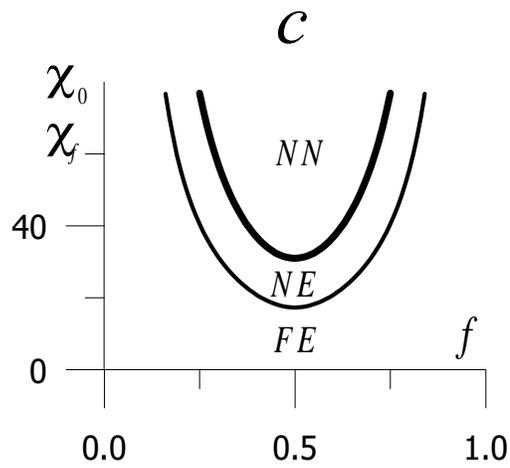
\item{Fig.7}   Phase diagram of the disordered heteropolymer:
a) at $\sigma = 0$, $a = 1$; $l = 0.1$;
b) at $\sigma = 0$, $a = 1$; $l = 5$;
c) at $\sigma = 2$, $a = 1$; $l = 0.1$.
Thick curve corresponds to the ergodicity breaking point and thin
curve corresponds to freezing point; $FE$, $NE$, $NN$ correspond to
freezing ergodic, non-freezing ergodic, non-freezing non-ergodic
phases.

\end{description}
\end{document}